\documentclass[%
 reprint,
superscriptaddress,
showpacs,
 amsmath,amssymb,
 aps,
]{revtex4-1}

\usepackage{relsize}
\usepackage{graphicx}
\usepackage{dcolumn}
\usepackage{bm}
\usepackage{color}
\usepackage{hyperref}

\begin{document}

\preprint{APS/123-QED}

\title{High-harmonic generation enhanced by dynamical electron correlation}

\author{Iliya Tikhomirov}

\affiliation{Department of Nuclear Engineering and Management, Graduate School of Engineering, The University of Tokyo, 7-3-1 Hongo, Bunkyo-ku, Tokyo 113-8656, Japan}

\author{Takeshi Sato}
\email{sato@atto.t.u-tokyo.ac.jp}
\affiliation{Department of Nuclear Engineering and Management, Graduate School of Engineering, The University of Tokyo, 7-3-1 Hongo, Bunkyo-ku, Tokyo 113-8656, Japan}
\affiliation{Photon Science Center, Graduate School of Engineering, The University of Tokyo, 7-3-1 Hongo, Bunkyo-ku, Tokyo 113-8656, Japan}

\author{Kenichi L. Ishikawa}
\email{ishiken@n.t.u-tokyo.ac.jp}
\affiliation{Department of Nuclear Engineering and Management, Graduate School of Engineering, The University of Tokyo, 7-3-1 Hongo, Bunkyo-ku, Tokyo 113-8656, Japan}
\affiliation{Photon Science Center, Graduate School of Engineering, The University of Tokyo, 7-3-1 Hongo, Bunkyo-ku, Tokyo 113-8656, Japan}

\date{\today}

\begin{abstract}
We theoretically study multielectron effects in high-harmonic generation (HHG), using all-electron first-principles simulations for a one-dimensional (1D) model atom. In addition to usual plateau and cutoff (from a cation in the present case, since the neutral is immediately ionized), we find a prominent resonance peak far above the plateau and a second plateau extended beyond the first cutoff. These features originate from the dication response enhanced by orders of magnitude due to the action of the Coulomb force from the rescattering electron, and, hence, are a clear manifestation of electron correlation. Although the present simulations are done in 1D, the physical mechanism underlying the dramatic enhancement is expected to hold also for three-dimensional real systems. This will provide new possibilities to explore dynamical electron correlation in intense laser fields using HHG, which is usually considered to be of single-electron nature in most cases.

\end{abstract}

\maketitle


 
Atoms and molecules interacting with intense ($\gtrsim 10^{14}\,{\rm　W/cm}^2$) visible-to-midinfrared laser pulses exhibit nonperturbative nonlinear response such as above-threshold ionization (ATI), tunneling ionization, high harmonic generation (HHG) and nonsequential double ionization (NSDI){\cite{Protopapas1997RPP}.
HHG, especially, forms the basis for attosecond science \cite{Agostini2004RPP,Krausz2009RMP,Gallmann2013ARPC} as highly successful means to generate attosecond coherent light pulses in the extreme-ultraviolet (XUV) and soft x-ray regions \cite{Chang,Zhao2012OL,Takahashi2013NatComm,Popmintchev2012Science} as well as to probe the electronic structure \cite{Itatani2004Nature,Smirnova2009Nature} and dynamics \cite{Calegari2014Science,Worner2015Science,Okino2015ScienceAdvanced} in atoms and molecules.  In the context of the latter, it is crucial to understand how HHG spectra reflect the electronic structure. As representative examples in atomic systems, the Cooper minimum in Ar \cite{Worner2009PRL}, autoionizing resonance in ${\rm Sn}^+$ \cite{1367-2630-15-1-013051}, and the giant resonance in Xe \cite{Pabst2013PRL} have been reported to imprint themselves in HHG spectra. All of these can be understood basically as features of single-photon ionization, i.e., the inverse process of recombination, which is the last step in the semiclassical three-step model \cite{Corkum1993PRL,Kulander1993} of HHG.
 
In this Letter, we predict a new mechanism leading to a drastic enhancement in HHG spectra, induced by the interaction of the recolliding electron with the electrons in the parent ion. We numerically simulate HHG from a one-dimensional (1D) multielectron model atom using a recently developed first-principles method called the time-dependent complete-active-space self-consistent-field (TD-CASSCF) method \cite{Sato2013PRA,Ishikawa2015JSTQE,Sato2016PRA}. We find, in harmonic spectra from 1D Be, a prominent peak that cannot be attributed to any resonant transition in Be and ${\rm Be}^+$.   Our analyses reveal that, whereas the cation (${\rm Be}^+$) plays a dominant role in the formation of the main plateau (neutral Be is immediately ionized and does not contribute to HHG), the peak originates from resonant excitation in the dication (${\rm Be}^{2+}$) induced by the recolliding electron. In addition, the action of the rescattering electron also drastically enhance harmonic generation from the dication to form the second plateau in the HHG spectrum. Thus, HHG spectra can reflect not only the electronic structure of the species (typically a neutral atom but a cation in this study) from which the returning electron is released \cite{Worner2009PRL,1367-2630-15-1-013051,Pabst2013PRL}, but also that of the parent ion (typically a cation but a dication in this study) with which the returning electron collides. The enhancement of the latter signal is a clear manifestation of multielectron effects.


The Hamiltonian for a 1D $N$-electron model atom interacting with an external laser electric field $ E(t) $ is taken in the length gauge as (we use the Hartree atomic units throughout unless otherwise stated),
\begin{align}
\label{eq:Hamiltonian}
	H =& \sum^{N}_{i=1}\left[- \frac{1}{2}\dfrac{\partial^2}{\partial x_{i}^{2}} -\frac{4}{\sqrt{x_{i}^2+1}} - E(t)x_{i}\right] \nonumber \\
	&+ \sum^{N}_{i>j} \frac{1}{\sqrt{(x_{i}-x_{j})^2+1}} ,
\end{align}
where $x_i (i=1,\cdots,N)$ denotes the position of the $i$th electron, and we use soft Coulomb potentials for electron-nuclear and electron-electron Coulomb interactions. We simulate the electron dynamics governed by this Hamiltonian using the recently developed TD-CASSCF method \cite{Sato2013PRA,Ishikawa2015JSTQE,Sato2016PRA}. 

In this method, the $N$-electron wave function $\Psi (t)$ is expressed as a superposition,
\begin{equation}
	\Psi (t) = \sum_{I} C_I(t) \Phi_I (t), 
\end{equation}     
of Slater determinants $\Phi_I (t)$ built from a given number $n$ of orthonormal orbital functions $\{ \psi_p (t)\}$. Both configuration-interaction (CI) coefficients $C_I(t)$ and orbital functions are time-dependent, which allows the use of considerably fewer orbitals than in fixed-orbital approaches. The orbitals are flexibly classified into {\it core} and {\it active} subspaces (see Fig.~1 of Ref.~\cite{Sato2013PRA}). We assume that $n_C$ core orbitals, accommodating tightly bound electrons, are doubly occupied all the time, whereas we consider all the possible distributions of $N_A (=N-2n_C)$ electrons among $n_A$ active orbitals, to take account of strong excitation and ionization.

It is also possible to further split the core space into {\it frozen core} (FC, fixed with no response to the field) and {\it dynamical core} (DC, allowed to vary in time and respond to the field).
 Let us use notation $(n_{FC},n_{DC},n_A)$ for the TD-CASSCF method with $n_{FC}$ FC orbitals, $n_{DC}$ DC orbitals, and $n_A$ active orbitals. Note that $(0, 0, n)$ is equivalent to the multiconfiguration time-dependent Hartree-Fock (MCTDHF) method \cite{PhysRevA.71.012712, Kato2004533, Sawada2016PRA} with $n$ occupied orbitals. The equations of motion for CI coefficients and orbital functions are derived, based on the time-dependent variational principle \cite{Frenkel1934, LOWDIN19721, QUA:QUA560070414}. See Ref. \cite{Sato2013PRA} for detailed description of the TD-CASSCF method and Ref. \cite{Ishikawa2015JSTQE} for a broad review of ab initio methods for multielectron dynamics.    
	

In this study, we specifically consider a 1D Be model atom ($N=4$) and a laser field with a $\sin^2$ envelope,
\begin{equation}
	E(t)={E_0}\sin^2\frac{\pi t}{T}\sin \omega t \qquad (0\leq t \leq T),
\end{equation}
where $T$ denotes a foot-to-foot pulse width. For all the results presented in this work except for Fig.~\ref{fig:Graph1}(b), we use 750 nm central wavelength, $5.2\times 10^{14}\,{\rm W/cm}^2$ peak intensity, and 22 optical-cycle foot-to-foot pulse width. The peak intensity is
assumed to be $5.2\times 10^{14}\,{\rm W/cm}^2$. 
Orbital functions are discretized on 2,048 equidistant grid points with box size $|x|<400$. 
We implement an absorbing boundary by a $\cos^{1/4}$-shape mask function at 15 \% side edges of the box. The time propagation is performed using 10,000 time steps per optical cycle. 
 We have also tested up to four times smaller grid spacing and six times smaller time step, and confirmed that the results remain virtually the same. The initial ground state is obtained through imaginary-time propagation.
Harmonic spectra are calculated as the squared magnitude of Fourier transform of dipole acceleration. 
In order to reduce the background level of the spectra, 
simulations are continued for some periods after the end of the
pulse, and the dipole acceleration after the pulse is multiplied by a
polynomial damping function.

\begin{table}[tb]
	\caption{\label{tab:table1}%
		Ionization potential $I_p$ (evaluated through Koopmans' theorem), cutoff energy $E_c$, and barrier suppression intensity $I_{BS}$ of each species.
	}
	\begin{ruledtabular}
		\begin{tabular}{lccc}
			  & $I_p$ (eV) & $E_c$ (eV) & $I_{BS}$ ($ {\rm W}/{\rm cm}^{2} $) \\
			\colrule
			Be & 8.5 & 95.1 & $2.1\times 10^{13}$\\
			${\rm Be}^{+}$ & 22.5 & 109.1 & $2.56\times 10^{14}$\\
			${\rm Be}^{2+}$ & 65.4 &  152.0 & $8.12\times 10^{15}$\\
		\end{tabular}
	\end{ruledtabular}
\end{table}


Figure \ref{fig:Graph0} shows HHG spectra calculated with three different subspace decompositions schematically depicted in Fig.~\ref{fig:2}. We assume that $(0,0,10)$ is the most accurate. The resulting spectrum (red thick curve in Fig.~\ref{fig:Graph0}) exhibits two remarkable features:
\begin{enumerate}
	\item a second plateau and cutoff around 150 eV beyond the first cutoff around 110 eV
	\item a prominent peak at 30.6 eV, ca. $10^3$ times higher than the plateau
\end{enumerate}
Both are present also in the result for $(0,1,9)$ (black curve), whereas missing in that for $(1,0,9)$ (blue curve), which unambiguously indicates an essential contribution from the core electrons. 

In Table \ref{tab:table1} we list, for ${\rm Be}, {\rm Be}^+$, and ${\rm Be}^{2+}$, the ionization potential $I_p$ evaluated through Koopmans' theorem, the cutoff energy $E_c$ by a common formula $E_c = I_p + 3.17 U_p$ with $U_p$ being the ponderomotive energy, and the barrier suppression intensity $I_{BS}$ \cite{1367-2630-15-1-013051, doi:10.1142/S0218863595000343}. From the values of $E_c$ in this table, one notices that the first and second plateaus in Fig.~\ref{fig:Graph0} are due to HHG from the cation and dication, respectively. The neutral Be is immediately ionized and, thus, does not contribute to high-harmonic spectrum, since its barrier suppression intensity ($2.1\times 10^{13}{\rm W}/{\rm cm}^{2}$) is much smaller than the laser peak intensity~\cite{Ilkov92JoPB}.

In order to further confirm this, we separate the total harmonic spectrum into contributions from different charge states. For this purpose we calculate the contribution from the charge state $q+$ to dipole acceleration, or charge-state-resolved dipole acceleration $\ddot{d}_q$, conveniently defined as (cf. Appendix of Ref.~\cite{Sato2013PRA}), 
\begin{align}
	\label{eq:charge-state-resolved-dipole-acceleration}
	\ddot{d}_q (t) &\equiv 
	\left(\begin{array}{c}N \\ q \end{array}\right) \int_ >dx_1\cdots \int_> dx_q \int_< dx_{q+1}\cdots \int_< dx_N \nonumber\\
	&\times \Psi^*(x_1,\cdots,x_N,t) \,\ddot{x}\, \Psi(x_1,\cdots,x_N,t),
\end{align}
where $\int_<$ ($\int_>$) denotes integration over a region $|x|<X_0$ ($|x|>X_0$) with $X_0 = 20$ a.u. in this study. The acceleration operator $\ddot{x}$ is evaluated as described in \cite{Sato2016PRA}. In this equation, we have omitted the summation with respect to spin variables for simplicity. Whereas the contributions from neutral Be and ${\rm Be}^{3+}$ are negligible, the first plateau is dominated by the contribution from ${\rm Be}^+$, and the second plateau is formed by the response of ${\rm Be}^{2+}$ (Fig.~\ref{fig:Graph3}).

\begin{figure}[tb]
\centering
\includegraphics[width=\linewidth]{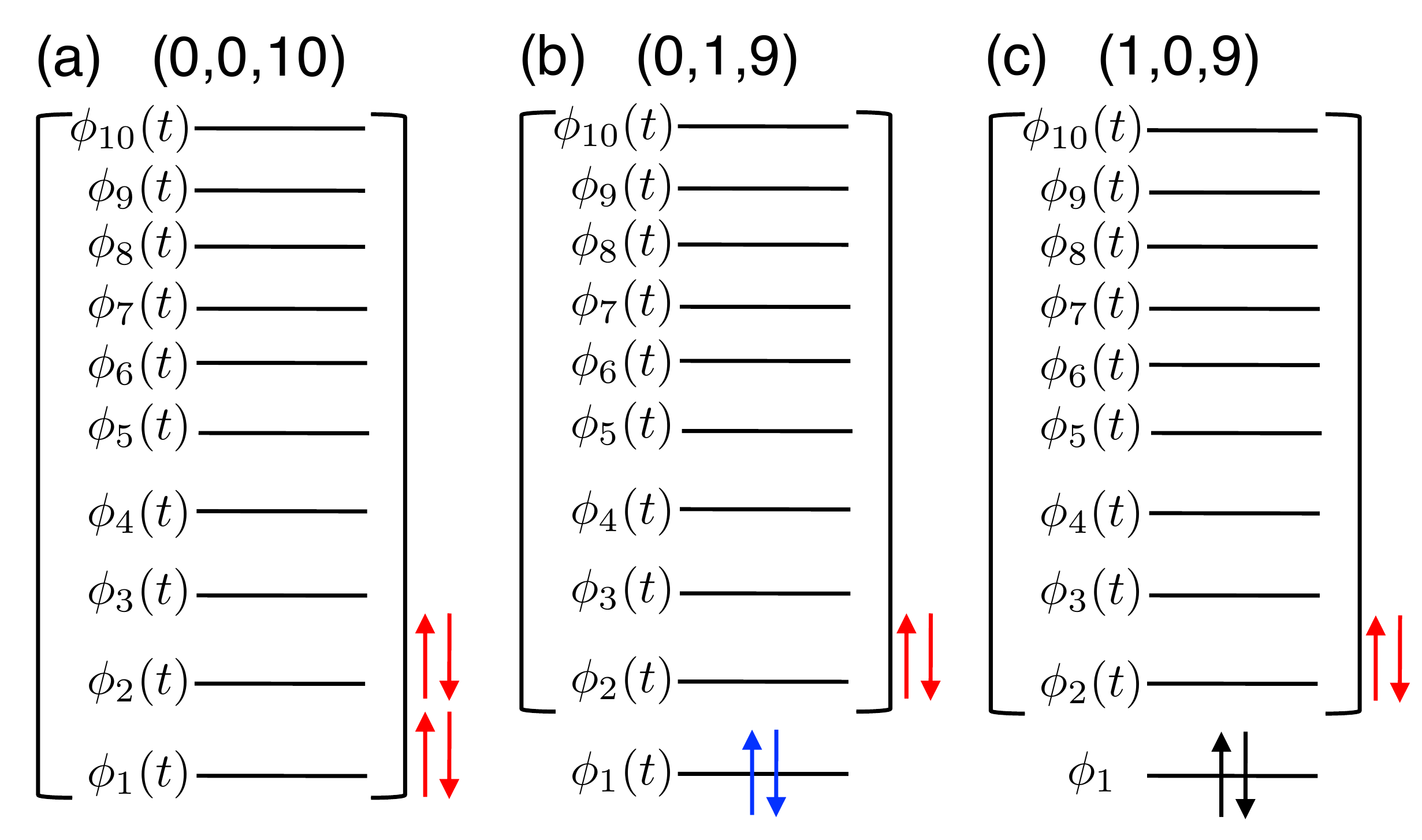}
\caption{(Color online) Pictorial explanation of orbital subspace decompositions for Be used in this study. (a) MCTDHF (0,0,10) with ten active orbitals, considered to be the most accurate (b) CASSCF (0,1,9) with one DC and nine active orbitals (c) CASSCF (1,0,9) with one FC and nine active orbitals.}
\label{fig:2}
\end{figure}

\begin{figure}[tb]
	\centering
	\includegraphics[width=1.\linewidth]{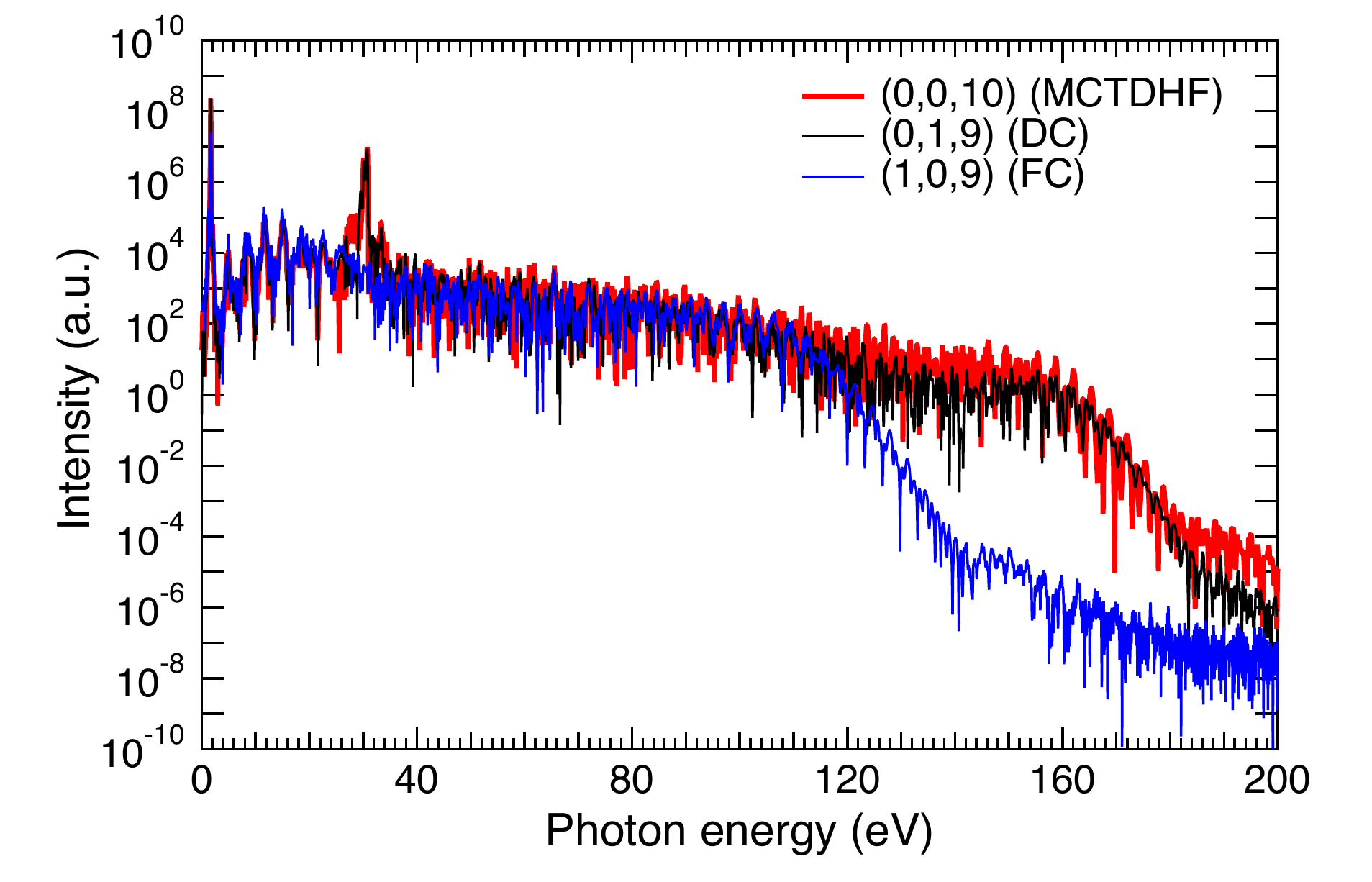}
	\caption{(Color online) High-harmonic spectra from a 1D Be model atom calculated with three different subspace decompositions depicted in Fig.~\ref{fig:2}}
	\label{fig:Graph0}
\end{figure}

\begin{figure}[tb]
	\centering
	\includegraphics[width=1\linewidth]{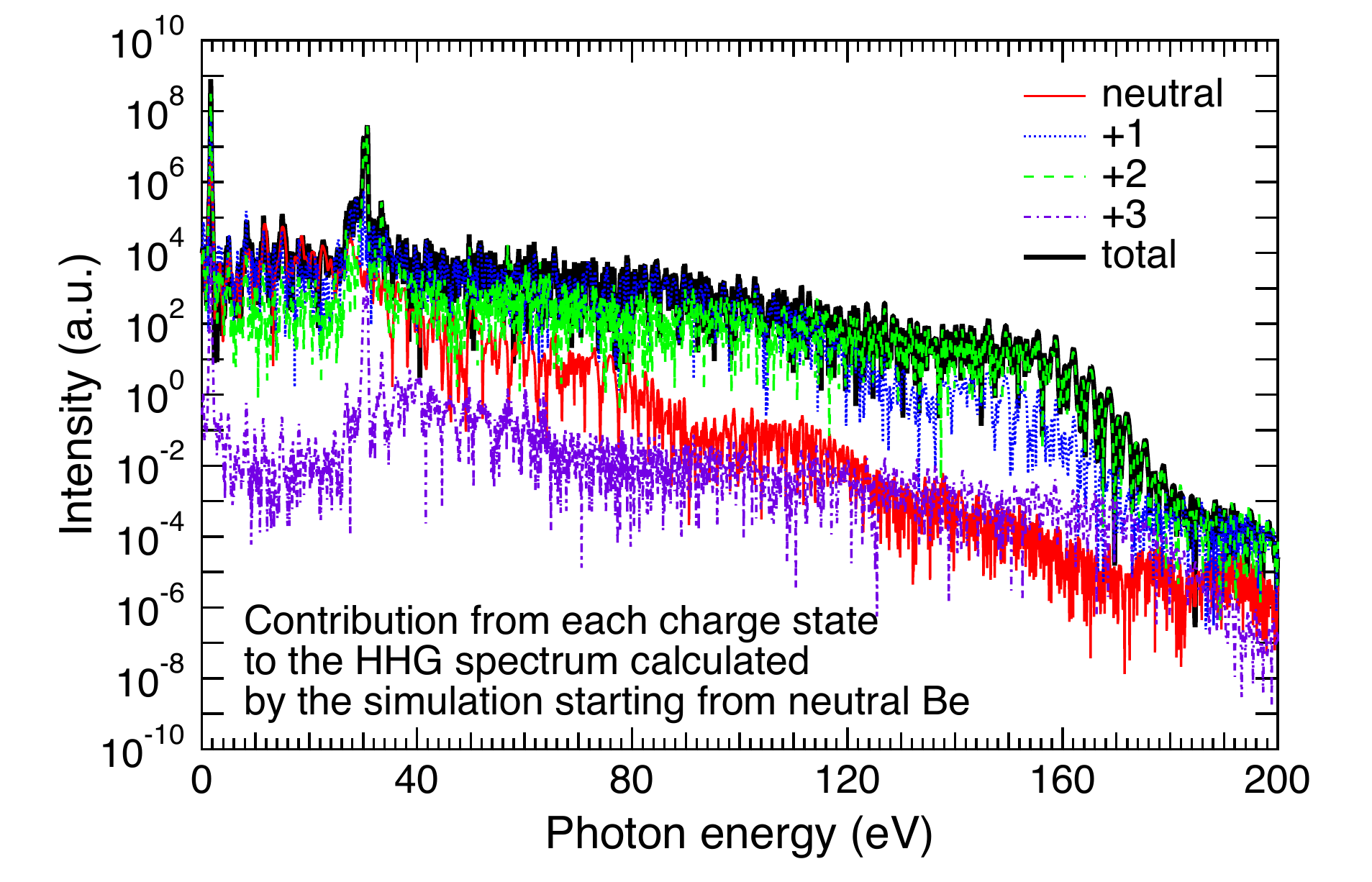}
	\caption{(Color online) Charge-state-resolved harmonic spectra extracted from the simulation starting from neutral Be and calculated as the squared magnitude of the Fourier transform of Eq.~(\ref{eq:charge-state-resolved-dipole-acceleration}). The total HHG spectrum (the same as the red curve in Fig.~\ref{fig:Graph0}) is also shown.} 
	\label{fig:Graph3}
\end{figure}

The sharp peak in Fig.~\ref{fig:Graph0} suggests the presence of an excitation resonance. Hence,  we compare the HHG spectra with
excitation spectra of Be, Be$^+$, and Be$^{2+}$ [Fig.~\ref{fig:Graph1}(b)], 
obtained by Fourier transform of the dipole response to a quasi-delta-function pulse with a field being finite at a single time step.
One can clearly see that the first excitation energy (30.6 eV) of ${\rm Be}^{2+}$ coincides with the peak position in the harmonic spectrum. Also in Fig.~\ref{fig:Graph3}, ${\rm Be}^{2+}$ predominantly contributes to the peak. Therefore, both of the two above-mentioned remarkable features originate from the response of ${\rm Be}^{2+}$.

In order to further investigate the contribution from different species separately, we show in Fig.~\ref{fig:Graph1}(a) HHG spectra calculated by use of the MCTDHF method with ten spatial orbitals} starting from Be, ${\rm Be}^+$, and ${\rm Be}^{2+}$, respectively. It should be noted that the curves for ${\rm Be}^+$ and ${\rm Be}^{2+}$ in Fig.~\ref{fig:Graph1}(a) are, unlike in Fig.~\ref{fig:Graph3}, the results of the simulations whose initial states are ${\rm Be}^+$ and ${\rm Be}^{2+}$, respectively. The spectra from Be and ${\rm Be}^+$ are similar, confirming that the neutral species, immediately ionized, do not contribute. On the other hand, if we start simulations from ${\rm Be}^{2+}$, no plateau but only up to the fifth harmonics are observed; this is in fact reasonable if we note that the laser intensity is much lower than the barrier suppression intensity (Table \ref{tab:table1}). In addition, although a peak at 30.6 eV, related to first excitation, can be seen, it is lower by many orders of magnitude than in the spectrum from ${\rm Be}^+$ and in Fig.~\ref{fig:Graph0}. These observations imply the existence of a mechanism to enhance the response of ${\rm Be}^{2+}$.

In order to explore the effect of a rescattering electron originating from ${\rm Be}^+$, we have simulated high-harmonic generation from ${\rm Be}^{2+}$ by adding a Coulomb field from the rescattering electron,
\begin{equation}
\label{eq:rsc-electron-field}
	H_{\rm rsc} = \sum_{i=1}^{N} \int \frac{\rho (x^\prime,t)}{\sqrt{(x_{i}-x^\prime)^2+1}}\,dx^\prime,
\end{equation} 
as an external field to the Hamiltonian for ${\rm Be}^{2+}$, where $N=2$, and $\rho (x,t)$ denotes the time-dependent probability density of the ${\rm Be}^+$ valence electron, which forms an oscillating dipole and is calculated in a separate simulation starting from ${\rm Be}^+$ with a frozen-core orbital $(1,0,1)$. The resulting spectrum is shown by the green dashed line in Fig.~\ref{fig:Graph1}(a). The comparison with the blue dotted line reveals that $H_{\rm rsc}$ dramatically enhances the harmonic response of ${\rm Be}^{2+}$, including the peak at 30.6 eV, and almost recovers that of ${\rm Be}^+$. Whereas this mechanism is similar to enhancement by an assisting harmonic pulse \cite{PhysRevLett.91.043002, PhysRevA.70.013412, PhysRevLett.99.053904, PhysRevA.80.011807}, the enhancement is due to direct Coulomb force from the oscillating dipole, rather than harmonics emitted from it. In the words of the semiclassical three-step model, the recolliding electron ejected from ${\rm Be}^+$ lifts ${\rm Be}^{2+}$ to the first excited state, which subsequently emits a photon to form the peak at 30.6 eV, and, at the same time, facilitate tunneling ionization, enhancing harmonic emission in the second plateau. The nonresonant components of the oscillating dipole also lead to virtual excitation and facilitate tunneling ionization of ${\rm Be}^{2+}$. Thus, electron-electron interaction plays an essential role in generation of the second plateau and the peak at 30.6 eV.

\begin{figure}[tb]
\centering
\includegraphics[width=1\linewidth]{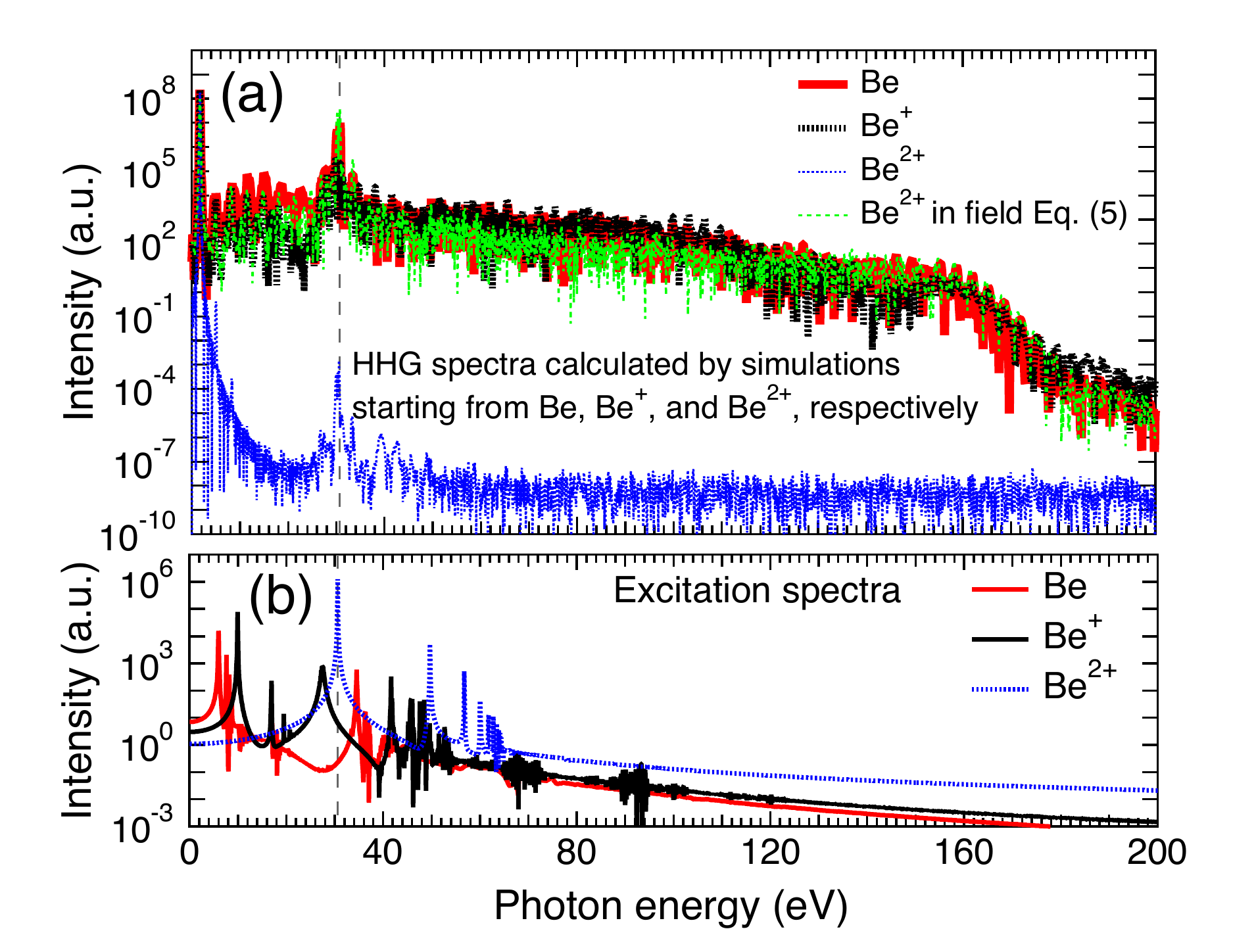}
\caption{(Color online) (a) High-harmonic spectra from 1D Be (thick solid red), ${\rm Be}^+$ (thick dotted black), and ${\rm Be}^{2+}$ (dotted blue), respectively, calculated with the MCTDHF methods with ten orbitals. The spectrum from Be$^{2+}$ in the field of the rescattering electron Eq.~(\ref{eq:rsc-electron-field}) is also shown (thin dashed green). Note that simulations have been started with Be, ${\rm Be}^+$, and ${\rm Be}^{2+}$, respectively, as an initial state. The thick solid red curve (Be) is the same as that in Fig.~\ref{fig:Graph0} and the thick solid black curve in Fig.~\ref{fig:Graph3}.
(b) Excitation spectra of 1D Be, ${\rm Be}^+$, and ${\rm Be}^{2+}$, respectively, calculated through excitation by a quasi-delta-function pulse. The highest peak is located at 30.6 eV and corresponds to the first excited state of 1D ${\rm Be}^{2+}$. The gray vertical dashed line indicates the position of 30.6 eV.}
\label{fig:Graph1}
\end{figure}

\begin{figure}[t]
	\centering
	\includegraphics[width=1\linewidth]{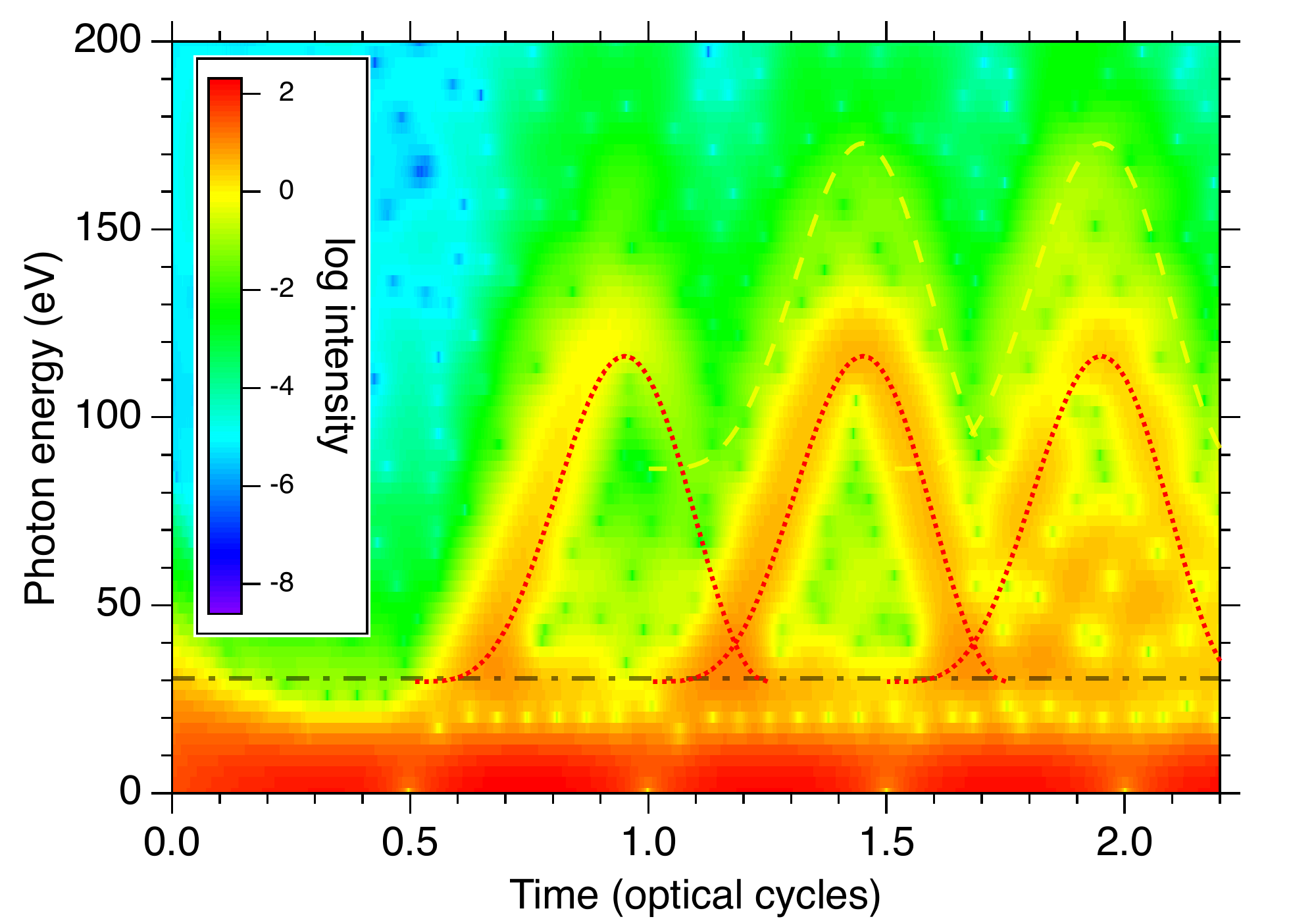}
	\caption{(Color online) Time-frequency analysis, or Gabor transform with a temporal window size of 0.095 cycle, of HHG from ${\rm Be}^+$ by a $5.2\times 10^{14}\,{\rm W/cm}^2$ flat-top pulse with a half-cycle ramp-on. Horizontal dash-dotted line: position of 30.6 eV. Dotted (dashed) curves: classically calculated kinetic energies of the returning electrons plus the ionization energy of ${\rm Be}^+$ (${\rm Be}^{2+}$). Note that the energy of quantum trajectories are slightly higher than that of classical ones due to finite distance between the origin and tunnel exit \cite{Lewenstein1994PRA,Ishikawa2010INTECH}.}
	\label{fig:time-frequency-analysis}
\end{figure}

It is expected from this scenario that photons at the peak are emitted at any time within an optical cycle while those in the first and second plateaus are emitted upon recombination of the rescattering electrons, the second plateau being delayed by a half cycle with respect to the first one. These are confirmed by Fig.~\ref{fig:time-frequency-analysis}, which shows the time-frequency analysis of HHG from ${\rm Be}^+$ by a $5.2\times 10^{14}\,{\rm W/cm}^2$ flat-top pulse with a half-cycle ramp-on. We can recognize typical arcs corresponding to the first (from ${\rm Be}^+$) and second (from ${\rm Be}^{2+}$) plateaus from the second and third half cycle, respectively. On the other hand, as expected, we see constant strong emission around 30 eV, which indicates that it is not due to recombination. Non-Born-Oppenheimer study on molecules \cite{Bandrauk2008PRL}, using the nuclear dynamics as a clock, may help disentangle these processes even more clearly.

The processes leading to the harmonic spectrum shown in Fig.~\ref{fig:Graph0} are summarized as follows. The laser intensity is so high that the neutral Be is completely depleted in the early stage of the pulse and hardly contributes to the spectrum. ${\rm Be}^+$ plays a dominant role in the formation of the first plateau and cutoff. Unexpectedly, the rescattering electron emitted from ${\rm Be}^+$ does not only contribute to the first plateau but also greatly enhances the response of ${\rm Be}^{2+}$, which leads to the formation of the sharp peak at 30.6 eV and the second plateau. It should be emphasized that, in contrast to resonance-induced enhancement mechanisms previously reported \cite{1367-2630-15-1-013051,Pabst2013PRL}, the peak is {\it not} related with the resonant excitation of ${\rm Be}^+$ but with that of ${\rm Be}^{2+}$.

In conclusion, we have investigated high-harmonic generation from 1D Be model atom using the all-electron TD-CASSCF method. In addition to the main plateau formed by HHG from ${\rm Be}^+$, we have found two prominent features: second plateau and a peak that is ca. $10^3$ times higher than the main plateau. Thanks to flexible subspace divisions characteristic of the TD-CASSCF, we have identified them as originating from ${\rm Be}^{2+}$. However, this response of ${\rm Be}^{2+}$ produced via tunneling ionization of Be and ${\rm Be}^+$ is totally different from that in the case where ${\rm Be}^{2+}$ is put under irradiation from the beginning. The recollision of the electron ejected from ${\rm Be}^{+}$ leads to dramatic enhancement of the response of ${\rm Be}^{2+}$ by exciting it and largely facilitating subsequent tunneling ionization. Thus, electron correlation plays an essential role in the appearance of the prominent peak and second plateau. Although we have specifically treated a 1D Be model atom, HHG enhancement by electron correlation as revealed here is presumably common to any 3D real target atoms and molecules. Whereas a resonance peak may be hidden under the main plateau, the enhancement of an extended plateau is expectedly more easily observable. The present study will open a new possibility to study multielectron effects and electron correlation using high-harmonic generation, which is usually considered a predominantly single-electron process.

We thank Armin Scrinzi for fruitful discussions.
This research is supported in part by Grant-in-Aid for Scientific
Research (No.~25286064, No.~26390076, No.~26600111, and No.~16H03881) from the Ministry of Education, Culture, Sports, Science and
Technology (MEXT) of Japan, and also by the Photon Frontier Network Program of MEXT. 
This research is also partially
supported by the Center of Innovation Program from the Japan Science and
Technology Agency, JST, and by CREST, JST.

\nocite{*}

\bibliography{apssamp}

\end{document}